\documentclass[usenatbib]{mn2e}

\usepackage[dvips]{graphicx}
\usepackage{amssymb}
\usepackage{txfonts}

\newcommand{\msun}{{\rm M}_{\sun}}
\newcommand{\xte}{{\textit{RXTE}}}

\title[Compton scattering in Cyg X-3]{Compton scattering as the explanation of the peculiar X-ray properties of Cyg X-3}

\author[A. A. Zdziarski, R. Misra and M. Gierli\'nski]
{Andrzej A.~Zdziarski,$^{1}$\thanks{E-mail: aaz@camk.edu.pl (AAZ), rmisra@iucaa.ernet.in (RM)} Ranjeev Misra$^2$\footnotemark[1] and 
Marek Gierli\'nski$^{3}$\\
$^{1}$ Centrum Astronomiczne im.\ M. Kopernika, Bartycka 18, 00-716 Warszawa, Poland\\
$^2$Inter University Centre for Astronomy and Astrophysics, Pune University Campus, Pune 411007, India\\
$^{3}$ Department of Physics, University of Durham, South Road, Durham DH1 3LE, UK\\
}

\date{Accepted 2009 October 27.  Received 2009 October 26; in original form 2009 May 7}

\pagerange{\pageref{firstpage}--\pageref{lastpage}}
\pubyear{2010}

\begin{document}

\maketitle

\label{firstpage}

\begin{abstract}
We consider implications of a possible presence of a Thomson-thick, low-temperature, plasma cloud surrounding the compact object in the binary system Cyg X-3. The presence of such a cloud was earlier inferred from the energy-independent orbital modulation of the X-ray flux and the lack of high frequencies in its power spectra. Here, we study the effect of Compton scattering by the cloud on the X-ray energy and power spectra, concentrating on the hard spectral state. The process reduces the energy of the high-energy break/cut-off in the energy spectra, which allows us to determine the Thomson optical depth. This, together with the observed cut-off in the power spectrum, determines the size of the plasma to be $\sim\! 2\times 10^9$ cm. At this size, the cloud will be in thermal equilibrium in the photon field of the X-ray source, which yields the cloud temperature of $\simeq 3$ keV, which refines the determination of the Thomson optical depth to $\sim\! 7$. At these parameters, thermal bremsstrahlung emission of the cloud becomes important as well. The physical origin of the cloud is likely to be collision of the very strong stellar wind of the companion Wolf-Rayet star with a small accretion disc formed by the wind accretion. Our model thus explains the peculiar X-ray energy and power spectra of Cyg X-3.
\end{abstract}

\section{Introduction}
\label{s:intro}

Cyg X-3 is a high-mass binary with a Wolf-Rayet companion \citep{v96}, with an unusually short orbital period of 4.8 h, located at a distance $d\simeq\! 9$ kpc in the Galactic plane \citep{d83,p00}. Due to the lack of reliable mass functions and determination of the inclination, it remains uncertain whether its compact object is a black hole or a neutron star (see \citealt{v09} for a recent discussion). However, the presence of a black hole is strongly favoured by considering the X-ray spectra, the radio emission, and the bolometric luminosity (\citealt{sz08}, hereafter SZ08; \citealt*{szm08}, hereafter SZM08; \citealt{h08}; \citealt{h09}, hereafter Hj09).

\begin{figure*}
\centerline{\includegraphics[width=6.7cm]{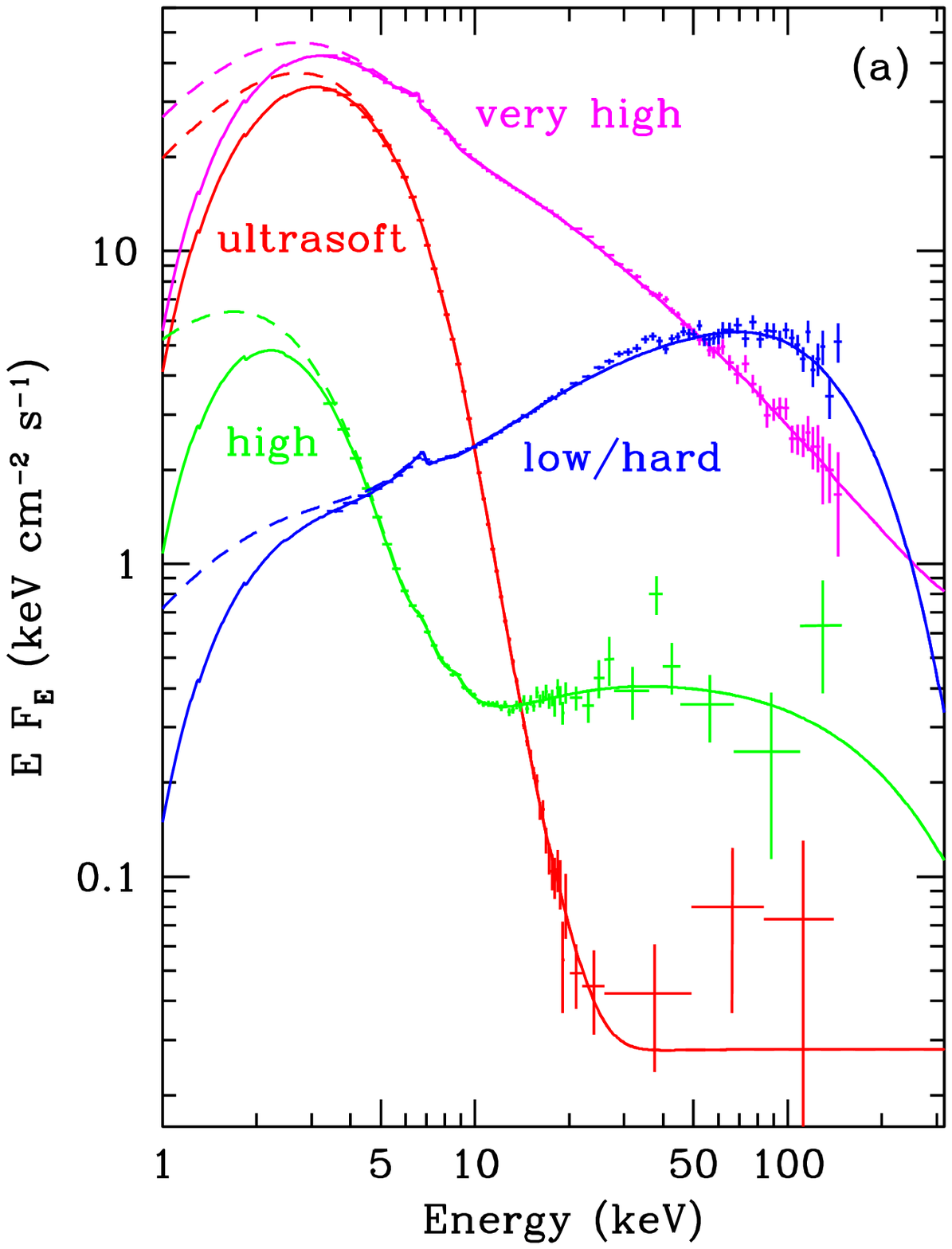}\hspace{0.5cm}
\includegraphics[width=7.5cm]{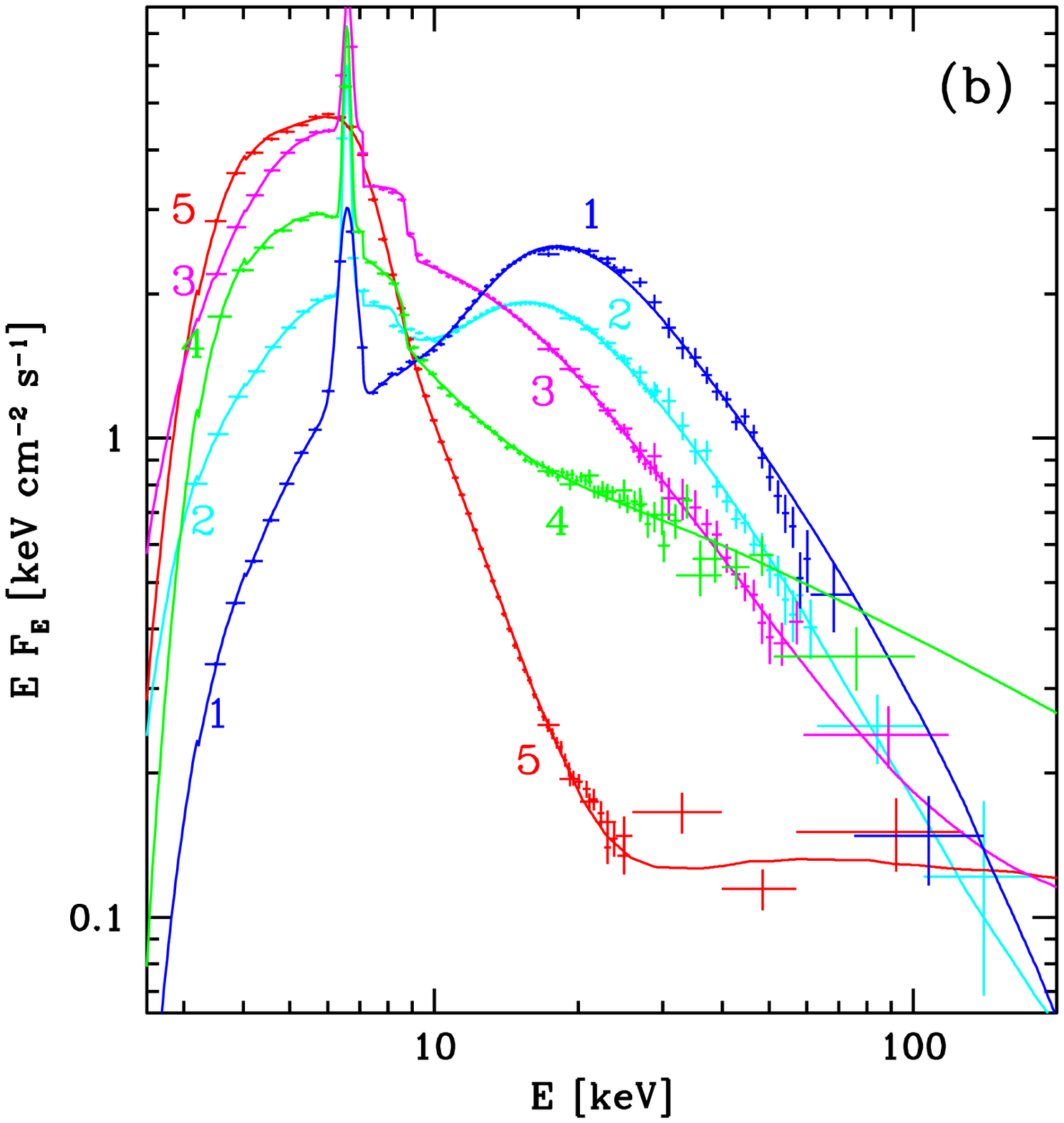}} 
\caption{(a) The canonical spectral states of black-hole binaries as exhibited by XTE J1550--564 \citep{zg04}. (b) The main spectral states of Cyg X-3 as classified by SZM08. Identical colours on both panels denote the corresponding states, except that both the blue (1) and cyan (2) spectra of Cyg X-3 correspond to the low/hard state. The spectra are modelled (solid curves) by hybrid (composite thermal and nonthermal) emission (\citealt{zg04}; SZM08); the dashed curves in (a) show the spectra before absorption.
} \label{1550_x3}
\end{figure*}

Cyg X-3 is a persistent X-ray source. Its X-ray spectra have been classified into 5 states by SZM08, who have also quantified their correlations with the radio states. Fig.\ \ref{1550_x3} compares the characteristic spectra of a typical black-hole binary, XTE J1550--564 \citep{zg04}, with the states of Cyg X-3. We see a remarkable overall similarity between the spectra of the two objects, apart from the very strong absorption in Cyg X-3. Its spectrum 5 is similar to the ultrasoft one of XTE J1550--564. Then, the spectra 3 and 4 are similar to the very high and high, respectively, states of XTE J1550--564. Finally, the hardest spectra of Cyg X-3, 1 and 2, can be identified with the canonical low/hard state of black-hole binaries. Here, however, the high-energy cut-off/break occurs in Cyg X-3 at a lower energy, $\sim\! 20$ keV, than in other black-hole binaries, where it is at $\sim\! 100$--200 keV, as seen in Fig.\ \ref{1550_x3}.

Initially, X-ray spectra of Cyg X-3 were modelled rather phenomenologically, e.g., by \citet{wh82}, \citet{n93}, \citet{r94}, including, e.g,, power-law, blackbody and/or bremsstrahlung components. More recently, physically motivated models have been studied by \citet{v03}, SZ08, \citet{h08} and Hj09. When the hard-state spectra were fitted by thermal Comptonization, the dominant process in the hard state \citep{zg04}, the obtained electron temperature was $kT_{\rm h}\sim\! 4$--8 keV, much lower than $kT_{\rm h}\sim\! 50$--100 keV in the usual hard state of black-hole binaries (e.g., \citealt{g97,z98,w02}). 

An important, but not included yet in the spectral studies, physical process in Cyg X-3 may be Compton scattering by a relatively cold, ionized, plasma with a Thomson optical depth $> 1$, surrounding the X-ray source in Cyg X-3. The presence of a highly ionized Thomson-thick absorber is indicated by a deep $\sim$9 keV Fe K edge seen in the X-ray spectra (see, e.g., SZ08). The presence of a Thomson-thick plasma in Cyg X-3 was postulated by, e.g., \citet{do74} and \citet*{hjr78} to account for the X-ray orbital modulation. Also, \citet{bv94} proposed the presence of such plasma based on the power spectra of Cyg X-3. That study, however, was hampered by an unresolved (at that time) instrumental effect, which prevented \citet{bv94} from performing a more detailed study of the effect of scattering on the power spectra. The lack or weakness of X-ray variability at frequencies of $f\ga 0.1$ Hz in Cyg X-3 was confirmed by \citet*{alh09}. We note that neither \citet{do74}, \citet{hjr78} nor \citet{bv94} discussed effects of the scattering cloud on the X-ray spectra of the source. This effect was mentioned by Hj09, but not taken into account in their fits due to the lack of a readily available spectral model.  

The location of the Thomson-thick plasma remains unclear. \citet*{v92} and \citet{v93} estimated the Thomson optical depth of the stellar wind from the Wolf-Rayet companion measured from the compact object towards the observer to be $\sim\! 3$. On the other hand, SZ08 found it to be $<\! 1$. Generally, as discussed by, e.g., SZ08 or \citet{v96}, the estimates of the mass-loss rate and the wind clumping in Cyg X-3 are uncertain, making these estimates relatively uncertain. The wind may also form a circumbinary envelope \citep{v07}, which may be optically thick. On the other hand, outflows from an accretion disc around the compact object may be optically thick as well, e.g., \citet{p07}. However, such outflows are usually axially symmetric around the compact object, which would not result in the observed orbital flux modulation. Another possibility for the physical identification of the scattering cloud is a bulge formed by the stellar wind colliding with the accretion disc. 

In this paper, we study in detail the effect of the presence of such plasma surrounding the X-ray source, on both the X-ray spectra (Section \ref{s:spectra}) and on the power spectra (Section \ref{s:timing}) in the hard state. In Section \ref{s:equilibrium}, we present our final model reproducing the energy and power spectra of Cyg X-3 and satisfying thermal equilibrium in the scattering cloud. In Section \ref{s:discussion}, we discuss our results and possible application of our model to the soft states of Cyg X-3.

\section{Compton down-scattering of X-rays}
\label{s:spectra}

\begin{figure}
\centerline{\includegraphics[width=7.cm]{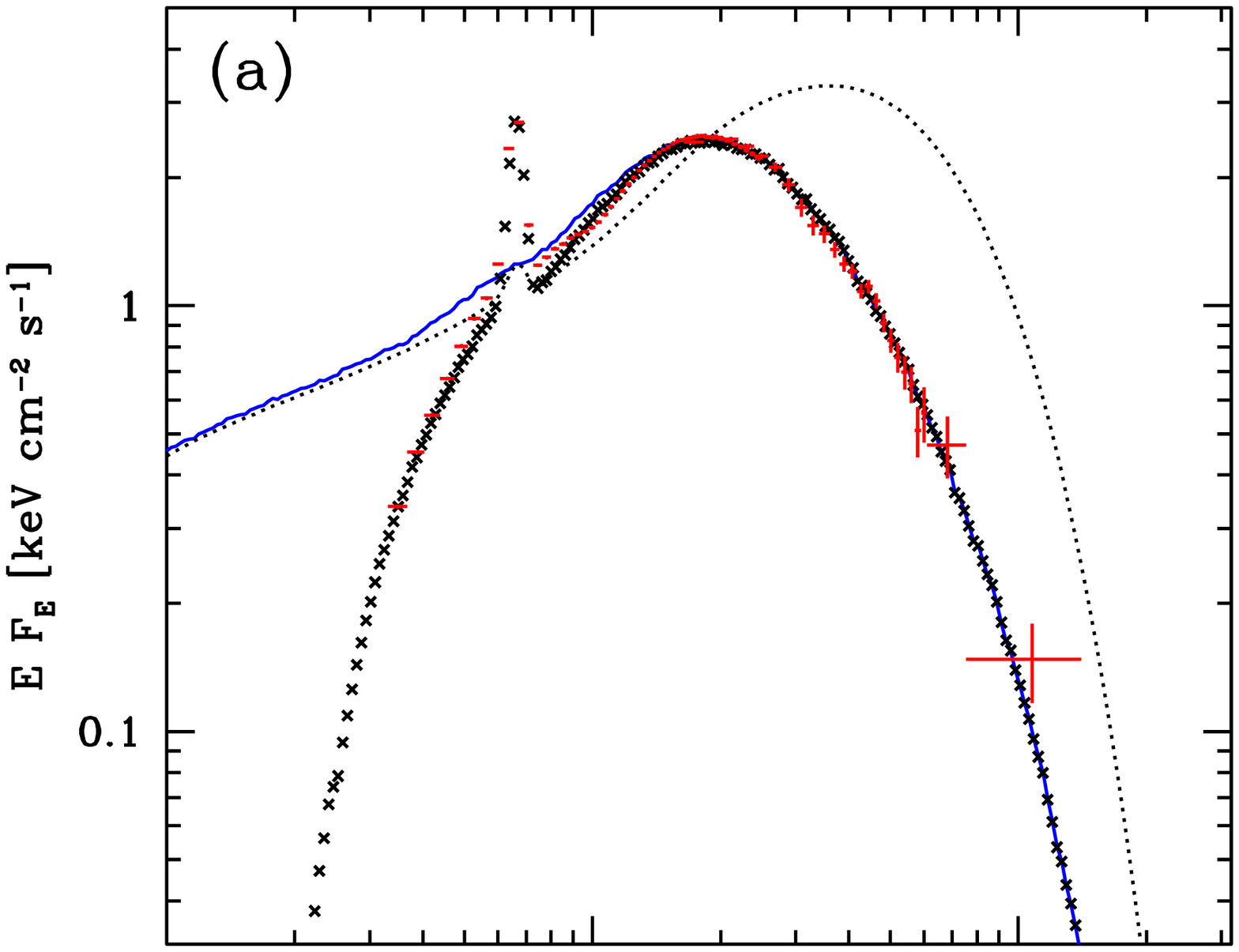}} \centerline{\includegraphics[width=7.cm]{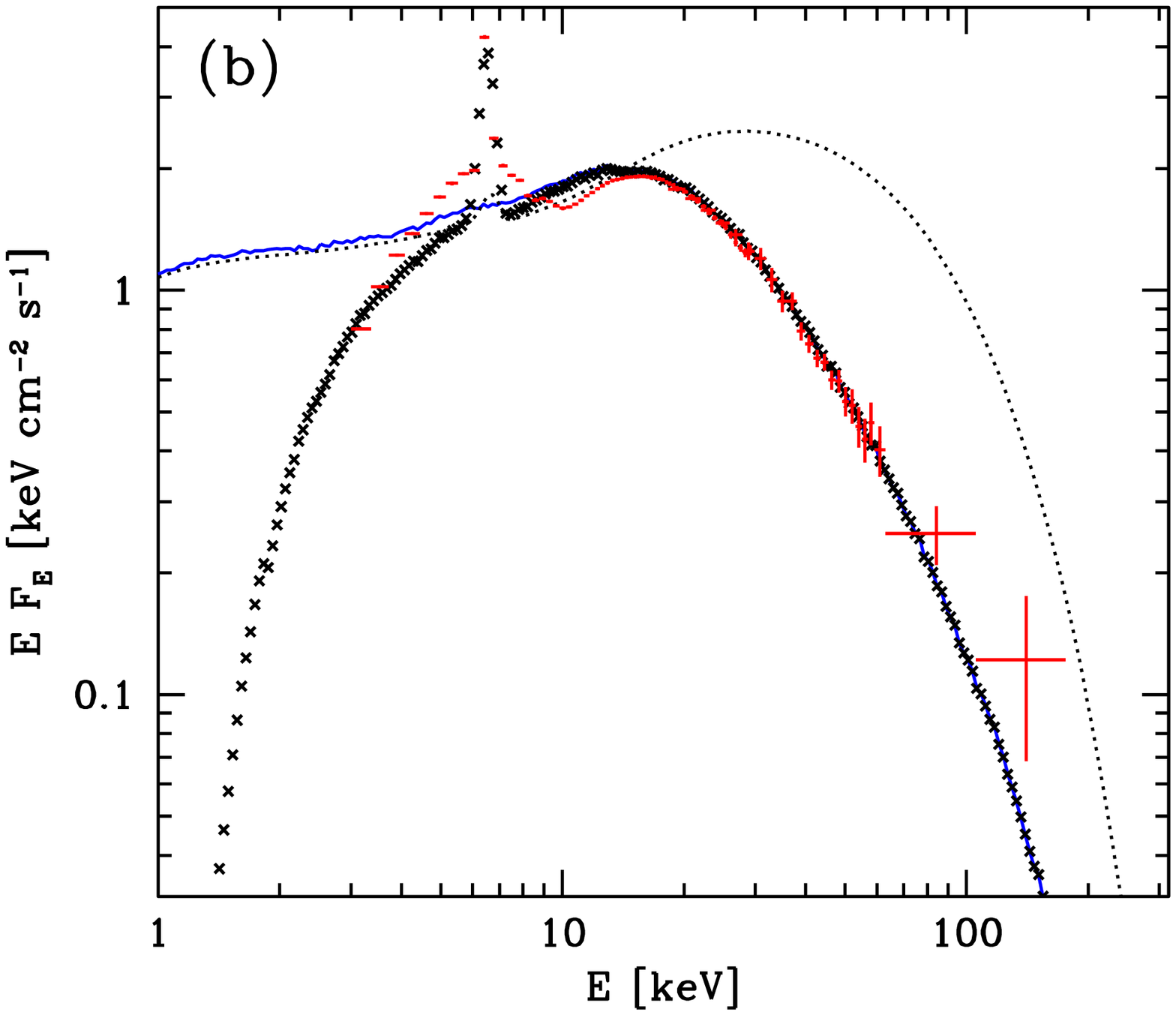}}
\caption{Effect of Compton down-scattering on X-ray spectra in Cyg X-3. The error bars show the average hard-state spectra, groups 1 and 2 of SZM08, the dotted curves show the incident spectra, and the solid curves show the spectra after transmission through a sphere of cold ionized electrons with the optical depth of $\tau_0$. The incident spectra are from thermal Comptonization in a hot plasma with an electron temperature, $kT_{\rm h}$, and the Compton parameter, $y_{\rm h}$, and including Compton reflection corresponding to a solid angle of an ionized reflector of $2\upi$ with an associated Fe K line. The crosses show the transmitted spectra after absorption by an external matter with a column density, $N_{\rm H}$, and including an externally produced Fe K line. (a) The average spectrum 1 compared to the down-scattered thermal Compton spectrum with $kT_{\rm h}=17$ keV, $y_{\rm h}=1.4$, $\tau_0=4.5$, $N_{\rm H}=9\times 10^{22}$ cm$^{-2}$. (b) The spectrum 2 modelled by $kT_{\rm h}=25$ keV, $y_{\rm h}=0.9$, $\tau_0=5.0$, $N_{\rm H}=3.5\times 10^{22}$ cm$^{-2}$. 
}
\label{downscattering}
\end{figure}

We consider the average hard-state spectra 1 and 2 of SZM08 from the \xte\/ Proportional Counter Array (PCA) and the High Energy X-Ray Transient Experiment (for the log of the observations, see Hj09; note that those authors used the term 'intermediate' for the spectrum 2). The advantage of choosing the hard state is that it has usually a standard form with a pronounced high-energy cut-off, and it is well modelled by thermal Comptonization in a hot plasma (e.g., \citealt{g97,z98,w02,zg04}). On the other hand, high-energy tails in soft states exhibit a variety of spectral shapes, see, e.g., \citet{zg04}. 

We use the thermal Comptonization model of \citet{ps96} in spherical geometry (with the geometry parameter $= 0$), {\tt compps} in {\sc xspec} \citep{arnaud96}. As the parameters of the model we use the electron temperature, $T_{\rm h}$, and the Compton parameter, $y_{\rm h}=4\tau_{\rm h} kT_{\rm h}/m_{\rm e} c^2$, where $\tau_{\rm h}$ is the Thomson optical depth of the hot plasma, and $m_{\rm e}$ is the electron mass (which definition is suitable for $\tau_{\rm h}\la 1$). In addition, we include Compton reflection \citep{mz95} from a highly ionized medium with an associated Fe K line, assuming a reflector solid angle of $2\upi$ at an inclination of $60\degr$ and the line equivalent width of 100 eV. (Although we find below that the reflection features in the intrinsic spectra are strongly smeared out in the scattered spectra, we keep them for physical consistency.) We assume seed photons Comptonized by the hot plasma to have a disc blackbody spectrum with the maximum temperature of 0.3 keV. 

As shown e.g., by SZ08, intrinsic X-ray absorption in Cyg X-3 is very complex, caused by the stellar wind with at least two phases. Here, we do not attempt to reproduce this complexity as it is not related to the main goal of this paper. Instead, we simply model absorption (of the photons leaving the scattering cloud) as due to a neutral medium. Also, we do not physically model the observed Fe K line, produced apparently outside of the scattering cloud. Instead, we simply add a Gaussian line reproducing the observed line to our final spectra.

Since hard-state spectra of black-hole binaries extend well above 100 keV, we need to accurately take into account the Klein-Nishina effects in the postulated scattering medium. For that purpose, we use the Monte Carlo code of \citet{gw84} and \citet{g00}. The code has been extensively tested against other codes and solutions \citep*{wlz88,zjm96,zpj00,z03}. We use a spherical geometry with uniform electron density, $n_{\rm e}$, and the source of photons at the centre. We note that the scattering medium in Cyg X-3 is likely to be spherically asymmetric if it is formed by accretion, and it needs to be such in order to account for the observed orbital modulation (Section \ref{s:intro}). However, here we attempt to reproduce only the properties averaged over the orbit, in particular, the average X-ray spectra of SZM08, for which purpose the assumed spherical symmetry is fully sufficient. The input photons have the X-ray spectrum as described above. In this section, we consider only electrons at a temperature low enough for the effect of Compton up-scattering to be negligible. This happens when the Compton parameter of the scattering medium, $y_0\equiv 4 (kT_0/m_{\rm e}c^2)\tau_0^2$ (which definition is valid for $\tau_0\gg 1$), is $\ll 1$, where $T_0$ is the temperature, $\tau_0= n_{\rm e}\sigma_{\rm T} R$ is the Thomson optical depth, $R$ is the radius, all of the scattering cloud, and $\sigma_{\rm T}$ is the Thomson cross section. In this section, we assume $kT_0=0.2$ keV, which satisfies the above condition for the values of $\tau_0$ we find here.

The models we consider are relatively complex. In particular, they involve a convolution of intrinsic spectra produced by thermal Comptonization in a hot plasma with the effect of Compton scattering in the, much colder, surrounding cloud. In Section \ref{s:equilibrium} below, we also find varying of the cloud temperature to be important and Comptonization of thermal bremsstrahlung to be a major process. At present, convolution models describing Compton scattering in a plasma of arbitrary Thomson optical depth are not available in {\sc xspec}. Given the complexity of this problem, we defer obtaining Green's functions for Comptonization to a future work. Thus, we are unable to formally fit X-ray count spectra. Instead, we compare our models to the X-ray spectra of SZM08, which were de-convolved assuming hybrid-plasma models. This will only enhance some discrepancies of the present models with the spectra of SZM08 obtained using different models, but it would not lead to a false agreement.

We have found that we can reproduce the average hard-state spectra with $\tau_0\simeq 5$, $kT_{\rm h}\sim 20$ keV and $y_{\rm h}\sim 1$, see Fig.\ \ref{downscattering} for details. The position of the peak of the down-scattered spectrum depends only weakly on $kT_{\rm h}$, as it is roughly at $m_{\rm e}c^2/\tau_0^2$ \citep{st80}; however, the slope of the high-energy tail of the spectrum depends sensitively on it. The hardness of the spectrum below the peak depends on the Compton parameter, $y_{\rm h}$; therefore we obtain a lower value of it for the softer spectrum. As seen in Fig.\ \ref{downscattering}, we can reproduce rather well the high-energy parts of the spectra, $\ga\! 10$ keV. On the other hand, as discussed above, accurate modelling of the part of the spectrum at $\la\! 10$ keV would require taking into account complicated physical processes in the inhomogeneous stellar wind, which we do not attempt here. 

\section{The effect of Compton scattering on timing properties}
\label{s:timing}

\citet*{llr81} calculated the distribution of the number of scatterings experienced by a photon emitted at the centre of an optically-thick sphere of cold electrons, given by their equation (18). \citet{kk87} used that result to calculate the distribution of the time spent by a photon inside the sphere, ${\rm d}N/ {\rm d}t$, with certain additional approximations. In particular, they assumed that a photon scattering $n$ times has covered the Thomson optical depth of exactly $\tau=n$ along its path, i.e., they replaced directly $n$ by $\tau$ in equation (18) of \citet{llr81}. On the other hand, the distribution of the number of scatterings after a given elapsed time is Poissonian, see, e.g., equation (6) in \citet{i79}. Thus, we can sum over $n$ for a given elapsed time, improving the accuracy with respect to the results of \citet{kk87}. In particular, we find,
\begin{equation}
{{\rm d}N\over {\rm d}\tau}={2\upi^2\over 3\tau_0^2}  \exp\left[-\tau\left(1-{\rm e}^{-{\upi^2\over 3\tau_0^2}}\right)\right],
\label{new18a}
\end{equation}
for $\tau\ga 0.5\tau_0^2$. Here, $\tau$ is both the optical depth that a photon has travelled and the elapsed time in units of the Thomson time, $\tau = c t\tau_0/R$, where $R$ is the sphere radius.

We have compared these results against those from our Monte Carlo code in the spherical geometry. We have considered photon energies in the Thomson regime since power spectra measured by \xte\/ PCA correspond mainly to it. We note here that there is a difference between the distribution of the photon arrival time at the sphere surface and that measured by a remote observer. Apart from a constant time shift, the latter is equal to the distribution measured at a plane tangent to the sphere. In units of the Thomson time, the difference is $\Delta\tau=\tau_0(1-\cos\alpha)$, where $\alpha$ is the angle at which a photon leaves the sphere with respect to the radial direction. Thus, $\Delta\tau= 0$ for photons leaving the sphere radially, and $\tau_0$ for photons tangent to the sphere surface.

\begin{figure}
\centerline{\includegraphics[width=7.cm]{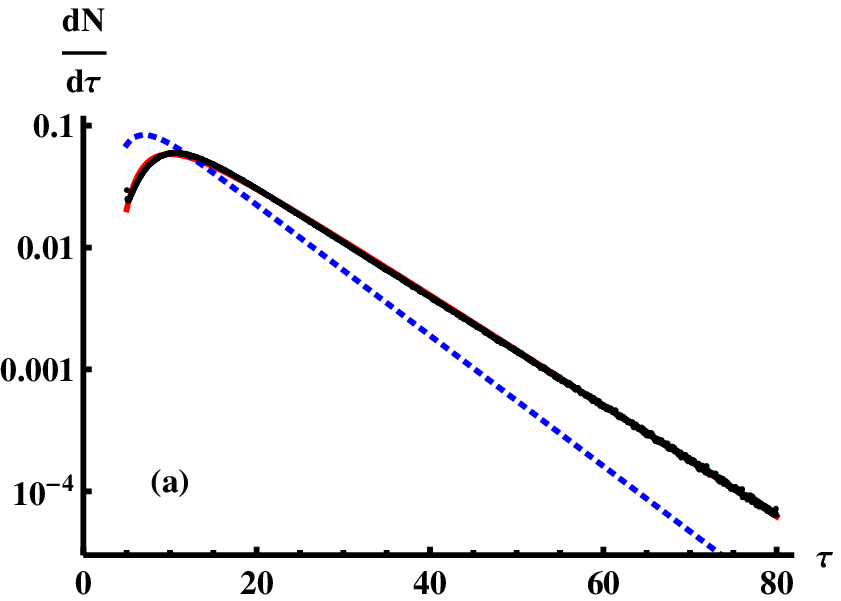}}
\centerline{\includegraphics[width=7.cm]{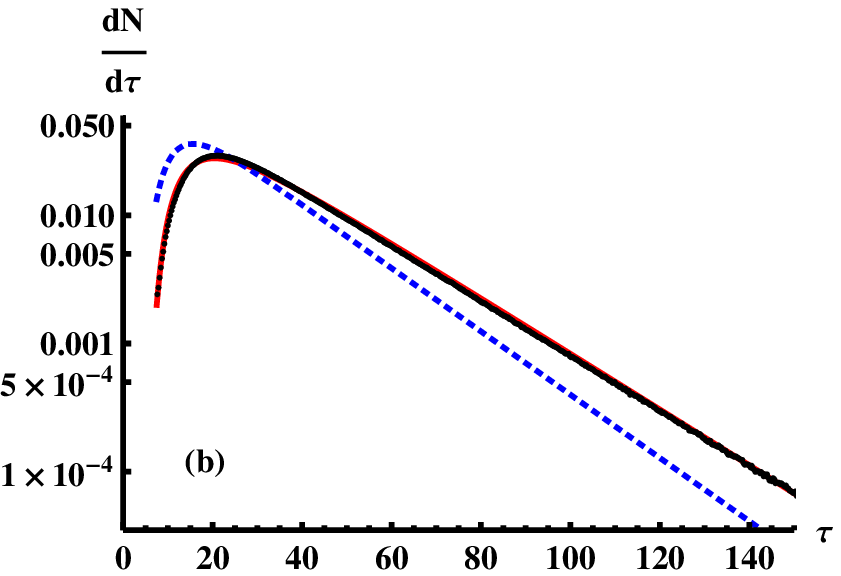}}
\caption{The distribution of the time spent by a soft photon in a spherical cloud of cold electrons measured from the cloud centre in units of the Thomson time for (a) $\tau_0=5$, (b) $\tau_0=7.5$. The black dots give Monte Carlo results and the red solid curve gives its fit, equation (\ref{fit}). The dotted curves gives the distribution of \citet{llr81} corrected for the Poissonian distribution of the time corresponding to a given number of scatterings. 
}
\label{tau5_7.5} 
\end{figure}

Using the Monte Carlo code, we have found that the results based on \citet{llr81} approximate the actual distribution only for a very large optical depth of the sphere, $\tau_0>20$ or so. Even including summation over the Poissonian distribution gives still a poor agreement with the Monte Carlo results at $\tau_0=5$, 7.5, as we show in Fig.\ \ref{tau5_7.5}. 

We have thus obtained a set of fitting formulae for ${\rm d}N/ {\rm d}\tau$ for values of $\tau_0$ of interest for this work. If normalized to unity, they also give the Green's functions, $G(\tau)={\rm d}N/ {\rm d}\tau$,
\begin{equation}
G(\tau)\simeq \cases{0, & $\tau<\tau_0$;\cr
r\exp\left[-{\tau\over \tau_1}-\left(\tau_2\over \tau\right)^2 \right]+\delta(\tau-\tau_0){\rm e}^{-\tau}, & $\tau\geq \tau_0$,\cr}
\label{fit}
\end{equation}
Here, $\tau_1$, $\tau_2$ are fitted, the term with the delta function accounts for the un-scattered photons, and $r$ follows from the normalization condition of $\int_0^\infty G(\tau){\rm d}\tau = 1$. For $\tau_0=5$ and 7.5, $\tau_1\simeq 9.513$, 19.87, $\tau_2\simeq 7.231$, 14.61, $r\simeq 0.2813$, 0.1299, respectively. We see in Fig.\ \ref{tau5_7.5} that equation (\ref{fit}) provides excellent fits to the Monte Carlo results. The actual distributions are significantly flatter than those corresponding to the results for $\tau_0\gg 1$. In particular, equation (\ref{new18a}) for $\tau_0=5$ yields $\tau_1\simeq 8.11$ (and $\tau_1=3\tau_0^2/\upi^2\simeq 7.60$ if the summation over the Poissonian distribution is neglected).

\begin{figure}
\centerline{\includegraphics[width=7.cm]{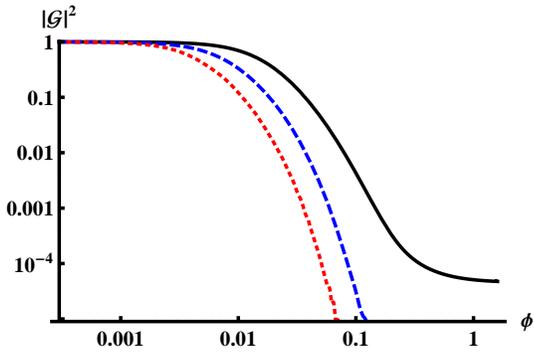}}
\caption{The squared attenuation coefficient as a function of frequency in the Thomson units ($\phi= f R/\tau_0 c$). The solid, dashed and dotted curves are for $\tau_0 =5$, 7.5, 10, respectively. At high frequencies, $|G(\phi)|\rightarrow {\rm e}^{-\tau_0}$. 
}
\label{attenuation} 
\end{figure}

If the intrinsic source variability is $S(\tau)$, the observed light curve, $F(\tau)$, is a convolution,
\begin{equation}
F(\tau)= \int_{\tau_0}^\infty S(\tau-\tau')  G(\tau') {\rm d}\tau'.
\label{convolution}
\end{equation}
The Fourier transform of $F(\tau)$, ${\cal F}$, is then the product of the individual transforms, i.e., 
\begin{equation}
{\cal F}(\phi)={\cal S}(\phi) {\cal G}(\phi), \qquad {\cal G}(\phi)=\int_{-\infty}^\infty G(\tau) {\rm e}^{-2\upi{\rm i}\phi \tau} {\rm d}\tau,
\label{product}
\end{equation}
where $\phi= f R /\tau_0 c$ is the dimensionless frequency in the Thomson units. If the original signal is sinusoidal, its amplitude is reduced\footnote{We note that we need to transform the full Green's function, including the un-scattered part, rather than to add ${\rm e}^{-\tau_0}$ to the absolute value of the transform of the scattered photons, which was done in \citet{kp89}.} by $\left|{\cal G}(\phi)\right|$. In general, damping due to scattering of an intrinsic variability can be calculated either using the convolution of the light curves, equation (\ref{convolution}), or by integrating the resulting power spectrum, equation (\ref{product}). The intrinsic power spectrum at $\phi$ is damped by $\left|{\cal G}(\phi)\right|^2$, which quantity is shown in Fig.\ \ref{attenuation}. We note that our Monte Carlo results show that the Fourier transform obtained by \citet{kp89}, which is based on the results of \citet{llr81}, becomes accurate only at $\tau_0\ga 20$.

Based on our Monte Carlo results, we have found that the characteristic dimensionless frequency, $\phi_{\rm c}$, which we define as that at which scattering reduces $\left|{\cal G}(\phi)\right|^2$ by a factor of e, is $\phi_{\rm c}\simeq 1/(2\tau_0^2)$, implying a characteristic dimensional frequency of $f_{\rm c}\simeq c/(2 R\tau_0)$. The characteristic frequency is 4 times lower than that obtained using the inverse of the (commonly used) average number of scattering in a sphere with a central source, $\tau_0^2/2$ (e.g., \citealt{st80}). Thus, our quantitative calculations demonstrate that Compton scattering is significantly more effective in damping variability than commonly assumed, affecting frequencies $\sim\! 4$ times lower than thought before. This is due to the relatively broad distribution of the number of photons per $\ln \tau$ ($=\tau{\rm d}N/{\rm d}\tau$).

\begin{figure}
\centerline{\includegraphics[width=7.cm]{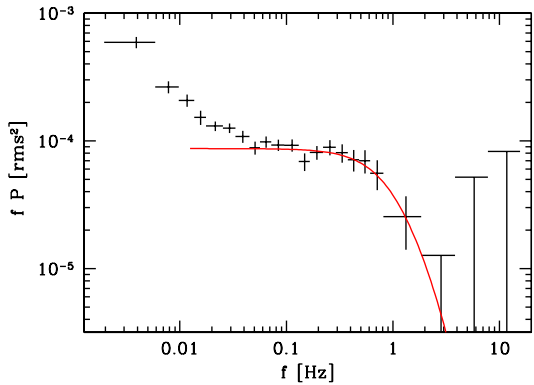}} \caption{The average power spectrum per $\ln f$ of Cyg X-3 from the \xte/PCA observations included in the average spectrum 1 (crosses). The solid curve shows a fit of $\left|{\cal G}(\phi)\right|^2$ for $\tau_0=7.5$ to the data at $f\geq 0.04$ Hz.
}
\label{power}
\end{figure}

We then compare our results with the observed power spectra of Cyg X-3. We have computed them for the hard-state \xte\/ observations used for the average hard-state spectrum 1 of SZM08, and the resulting average spectrum is shown in Fig.\ \ref{power}. We see a distinct cut-off at a frequency of $f_{\rm c}\simeq\! 1$ Hz. Using our results above, we can then deduce the size of the scattering cloud as 
\begin{equation}
R\simeq {c  \over 2 \tau_0 f_{\rm c}}\simeq {1.5\times 10^{10}\,{\rm cm} \over \tau_0 (f_{\rm c}/1\,{\rm Hz})}.
\label{size}
\end{equation}
For $f_{\rm c}\sim 1$ Hz and $\tau_0=5$, $R\sim 3\times 10^9$ cm, which is $\ll$ the separation between the components in Cyg X-3, $a\simeq 3\times 10^{11}(M/30\msun)^{1/3}$, where $M=M_{\rm X}+M_2$, and $M_{\rm X}$ and $M_2$ are the masses of the compact object and the donor, respectively. 

\section{Thermal equilibrium of the scattering cloud}
\label{s:equilibrium}

An important consequence of the relatively small size of the scattering cloud derived above is that its cooling by the companion star will be ineffective. The stellar luminosity is estimated as $L_*\simeq (1$--$30)\times 10^{38}$ erg s$^{-1}$ (see, e.g., a discussion in SZ08). Given $R/a\simeq 10^{-2}$, the cloud receives only $\sim\! 10^{-4}L_*$, which is $\ll$ the X-ray luminosity of $L_{\rm X}\sim 10^{38}$ erg s$^{-1}$ regardless of the exact value of $L_*$. Thus, the scattering cloud will be effectively heated by photons of the X-ray source and the surrounding accretion disc, and will achieve an equilibrium temperature, $kT_0$, with radiative gains balanced by radiative losses. The amplification factor, $A$, of the Comptonizing cloud can generally be written as,
\begin{equation}
{L_{\rm Ch}\over L_{\rm h}}= A(T_0,\tau_0,T_{\rm h},y_{\rm h}),
\label{ampli}
\end{equation}
where $L_{\rm h}$ is the luminosity of the central hot-plasma source (which is the only source of energy, and thus $L_{\rm h}=L_{\rm X}$), and $L_{\rm Ch}$ is the luminosity of this source after passing through the scattering cloud. For our parameters, $A<1$ usually (though not necessarily), depending on both the parameters of the cloud ($T_0$, $\tau_0$) and the shape of the intrinsic spectrum, determined by $T_{\rm h}$ and $y_{\rm h}$. 

We thus allow a free $kT_0$ imposing initially the condition of $L_{\rm Ch}=L_{\rm h}$. However, we find we can satisfy it only very roughly while keeping the model spectra fitting the data. The reason is that increasing $kT_0$ moves the peak in the down-scattered spectrum up, which requires an increase of $\tau_0$ (which moves the peak down) to compensate for it, which in turn reduces $L_{\rm Ch}/L_{\rm h}$, making it still $<1$. The obtained temperatures are $kT_0\sim 3$ keV, and the optical depths somewhat higher than in the models above, $\tau_0\simeq 7$. For these parameters, we find, however, that bremsstrahlung emission (e.g., \citealt{rl79}) of the scattering plasma is a substantial source of photons at the above $kT_0$ and $\tau_0$. The bremsstrahlung luminosity, $L_{\rm b}$, of a spherical cloud can be calculated to be,
\begin{eqnarray}
\lefteqn{
L_{\rm b}= {2^{7/2} \upi^{1/2}\alpha_{\rm f} \bar{g} q \mu_{\rm e} \tau_0^2 R m_{\rm e} c^3 \over 3^{3/2} \sigma_{\rm T}} \left(kT_0\over m_{\rm e} c^2\right)^{1/2}}
\label{brems}
\\
\lefteqn{\quad\,\, \simeq 1.8\times 10^{37} \bar{g} q \left(\tau_0\over 7.5\right)^2 {R\over 2\times 10^9\,{\rm cm}} {\mu_{\rm e}\over 2} \left(kT_0\over 3\,{\rm keV}\right)^{1/2}{\rm erg\,s}^{-1},}
\label{brems_num}
\end{eqnarray}
where $\mu_{\rm e}\simeq 2/(1+X)$ is the mean electron molecular weight, $X$ is the mass H relative abundance, $\alpha_{\rm f}$ is the fine-structure constant, $\bar g$ is the energy-averaged Gaunt factor, and $q$ accounts for 
a density gradient and/or clumpiness in the scattering cloud \citep*{ogs04},
\begin{equation}
q={\langle n_{\rm e}\rangle^2/ \langle n_{\rm e}^2\rangle}\geq 1.
\label{f}
\end{equation}
For consistency with our treatment of the timing properties, we consider a uniform cloud, where $q=1$. The factor of $\mu_{\rm e}$ follows from averaging over $Z^2$ at a given $\tau_0^2$ in a mixture of H and He, where $Z$ is the atomic charge. For cosmic composition, $X\simeq 0.7$ and $\mu_{\rm e}\simeq 1.2$. However, Cyg X-3 has a Wolf-Rayet companion, which composition is dominated by He (\citealt{v96}; see also a discussion in SZ08), for which $\mu_{\rm e}\simeq 2$. Thus, bremsstrahlung is, for a given $\tau_0$, a factor of $1+X\simeq 1.7$ more efficient in a He plasma than in a cosmic-composition medium. Note that $L_{\rm b}\propto R$ for a given $\tau_0$. We model the bremsstrahlung spectrum using the energy-dependent Gaunt factor of \citet*{kbk75}. The numerical values in equation (\ref{brems_num}) correspond to our final solution for the spectrum 1 below.

The bremsstrahlung emission is then Comptonized in the scattering cloud. Unlike Comptonization of the intrinsic central source, the seed photons are now uniformly distributed in the cloud (which follows from our assumption of the uniform cloud density), which we take into account in our Monte Carlo modelling. (We note that bremsstrahlung Comptonization could also have been dealt with using the Kompaneets equation, though less accurately than by the Monte Carlo method given the uniform distribution of the seed photons, see \citealt{st80}.) Comptonization depends (weakly) on bremsstrahlung self-absorption, which provides an effective low-energy cut-off for the seed photons. For our parameters, the energy at which the optical depth for self-absorption becomes unity is $E_{\rm s}\simeq\ 0.3$ eV (calculated following \citealt{rl79}), which value we adopt. The amplification factor of the bremsstrahlung emission, $A_{\rm b}$, can be written then as
\begin{equation}
{L_{\rm Cb}\over L_{\rm b}}= A_{\rm b}(T_0,\tau_0,E_{\rm s})>1,
\label{ampli_b}
\end{equation}
where $L_{\rm Cb}$ is the luminosity of the Comptonized bremsstrahlung emission of the cloud, and the dependence of $E_{\rm s}$ is weak.

The Comptonized bremsstrahlung emission by the electrons is an important cooling mechanism for the considered problem. The energy equilibrium equation can be written as,
\begin{equation}
L_{\rm h}=L_{\rm Ch}+L_{\rm Cb}.
\label{equilibrium}
\end{equation}
The size of the cloud, $R$, in equation (\ref{brems}) is then determined from equation (\ref{size}), which makes $L_{\rm b}\propto \tau_0$. 

Fitting the observed spectra while at the same time satisfying equations (\ref{size}), (\ref{equilibrium}) is a complex non-linear problem. To facilitate understanding of the requirements on the solution, we discuss now the effects of varying various parameters. An increase of $T_0$ moves the position of the peak in the scattered spectrum to higher energies. Also, it increases all $L_{\rm b}$, $L_{\rm Cb}$ and $L_{\rm Ch}$. On the other hand, the normalization of the bremsstrahlung spectrum below its high-energy cut-off at $\sim\! kT_0$ decreases as $T_0^{-1/2}$. An increase of $\tau_0$ moves the peak of the scattered spectrum down, though this saturates when the peak reaches $\sim\! 3kT_0$. Then, it increases $L_{\rm Cb}$ and decreases $L_{\rm Ch}$. Also, it steepens the slope of the scattered spectrum beyond its peak. An increase of $T_{\rm h}$ reduces $L_{\rm Ch}$ and makes the high-energy tail harder. An increase of $y_{\rm h}$ makes the slope below the peak harder, which holds for both the initial intrinsic spectrum and the scattered one. Finally, increasing $N_{\rm H}$ hardens the low-energy part of the spectrum without affecting the energy balance. 

\begin{figure}
\centerline{\includegraphics[width=7.cm]{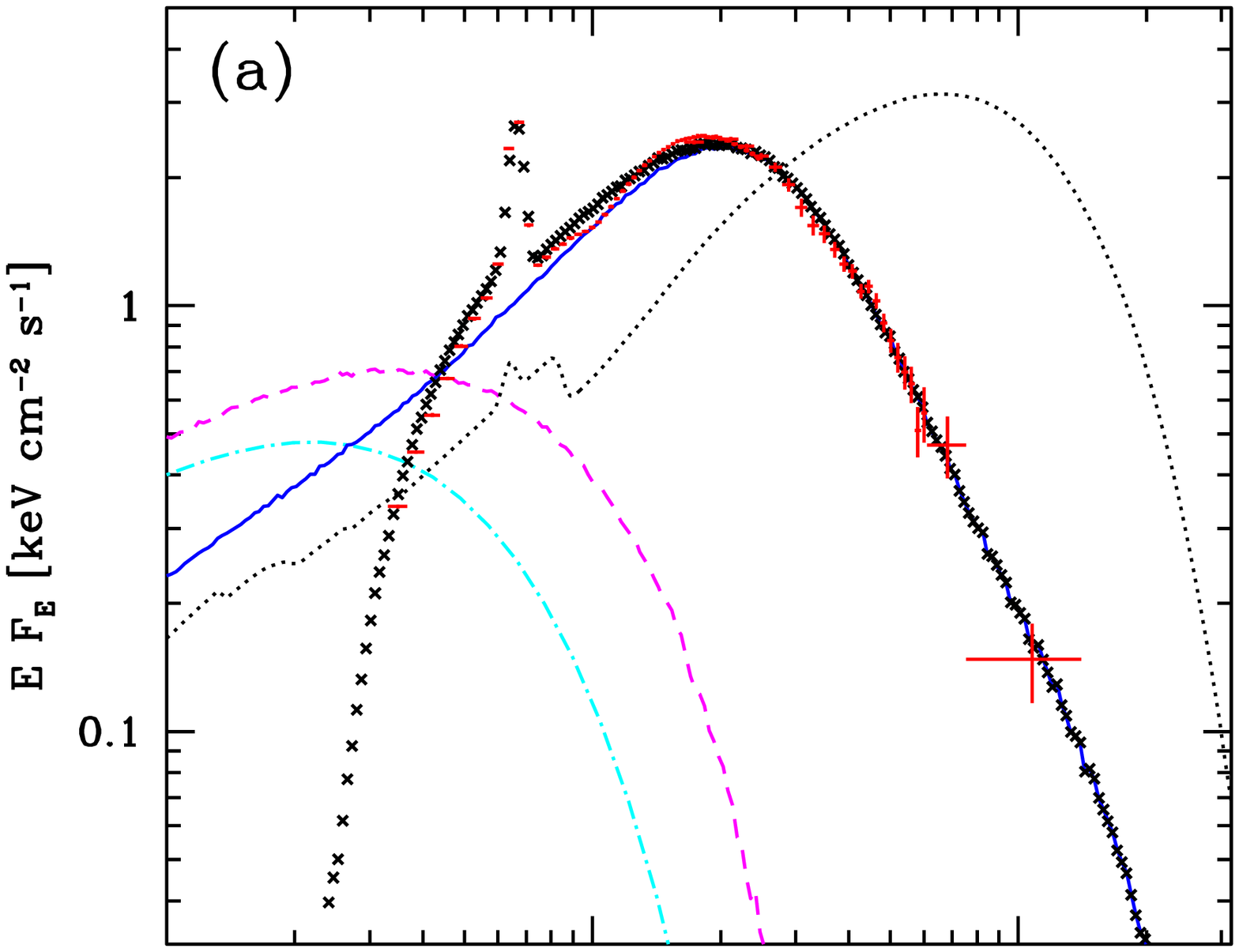}} 
\centerline{\includegraphics[width=7.cm]{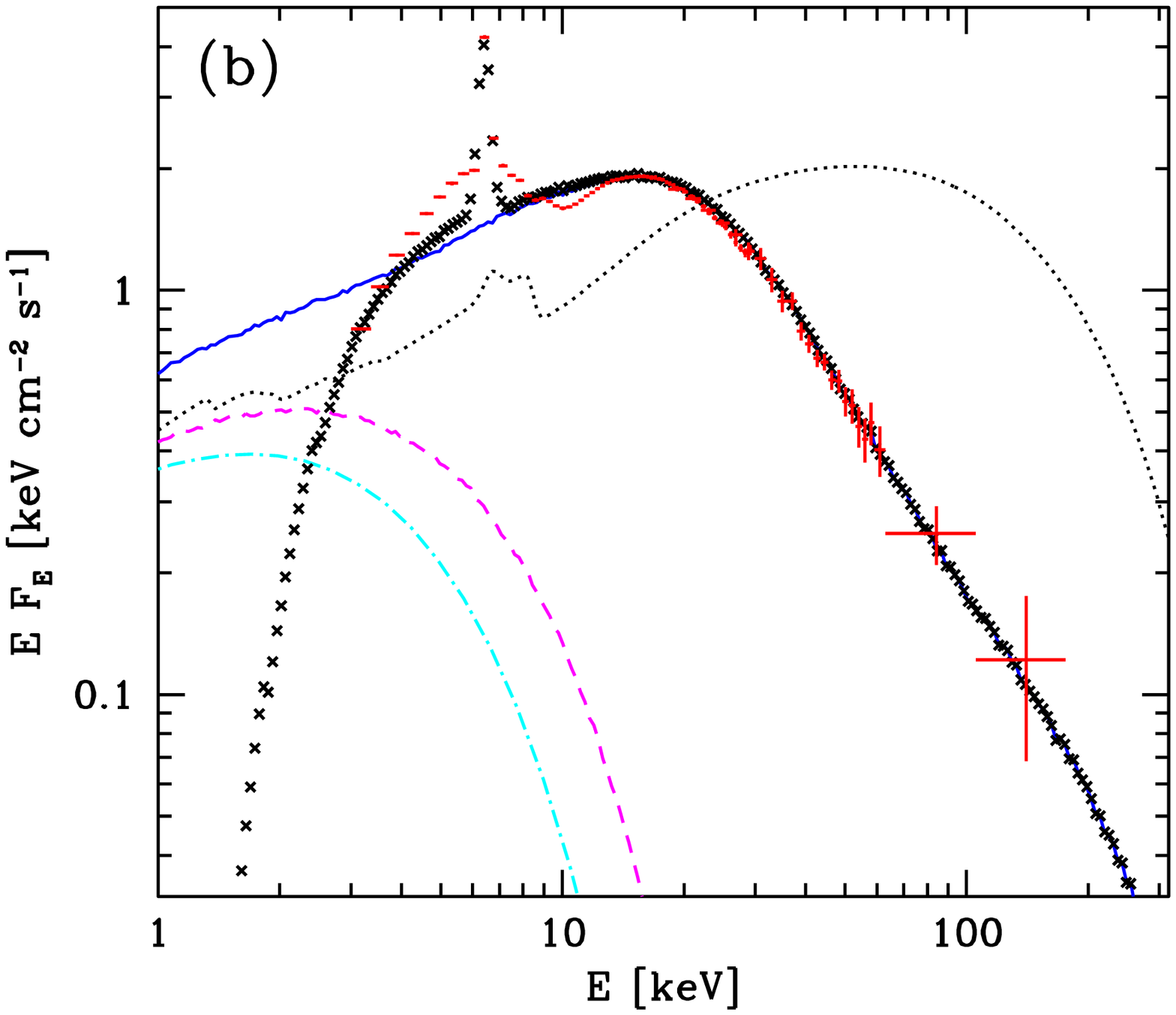}} 
\caption{The observed average spectra (error bars) compared to intrinsic thermal Compton spectra (dotted curves) transmitted (solid curves) through a Compton-thick cloud at the temperature, $T_0$, corresponding to thermal equilibrium and including the bremsstrahlung emission (dot-dashed curves) emitted by the cloud and Comptonized in it (dashed curves). The crosses show the sum of the Comptonized intrinsic and bremsstrahlung spectra after absorption by external matter with a column density, $N_{\rm H}$ and including an externally produced Fe K line. (a) The average spectrum 1 compared to the down-scattered thermal Compton spectrum with $kT_{\rm h}=30$ keV, $y_{\rm h}=2.0$, $\tau_0=7.5$, $kT_0=3.3$ keV, $N_{\rm H}=1.3\times 10^{23}$ cm$^{-2}$. (b) The spectrum 2 modelled by $kT_{\rm h}=50$ keV, $y_{\rm h}=1.2$, $\tau_0=7.0$, $kT_0=2.5$ keV, $N_{\rm H}=5\times 10^{22}$ cm$^{-2}$. In the thermal equilibrium, the unabsorbed intrinsic bolometric flux equals the sum of that flux after Compton scattering in the cloud and the flux of Comptonized bremsstrahlung. In both states, the bolometric flux is $1.1\times 10^{-8}$ erg cm$^{-2}$ s$^{-1}$.
}
\label{balance}
\end{figure}

We present the models reproducing the average spectra in Fig.\ \ref{balance}, which also gives the obtained parameters. We see in Fig.\ \ref{balance} that the spectrum 1 can be fitted very well. On the other hand, there is an indication of an additional component at low energies in the spectrum 2. Unfortunately, we cannot study it further with the \xte\/ data only, covering only the range $\ga 3$ keV. Still, the high energy parts are very well reproduced in both cases. The relative contribution of the Comptonized bremsstrahlung to the total luminosity is substantial, $\simeq 0.3$, 0.2 for the spectra 1 and 2, respectively. These factors also equal the fractional energy loss of the intrinsic emission in the scattering cloud. Comptonization amplifies the bremsstrahlung emission by factors $\simeq 1.5$, 1.3, respectively. We also find that including bremsstrahlung makes it possible to obtain good fits satisfying exactly the energy equilibrium, unlike the case without bremsstrahlung. The energy equilibrium is satisfied with 1 per cent in the cases shown in Fig.\ \ref{balance}. For the obtained parameters, the Compton temperature of the scattering cloud is $y_0\sim 1$. Then, photons at $E\ga 10$ keV are down-scattered, but photons at lower energies are efficiently up-scattered by the plasma. 

The obtained temperature of the hot Comptonizing plasma is now $kT_{\rm h}= 30$--50 keV. Generally, $kT_{\rm h}\sim 50$--100 keV are found in the hard state of black-hole binaries \citep{zg04}, with lower values found in more luminous states \citep{w02,y06,m08}\footnote{Note that temperatures as low as $\sim\! 20$ keV were claimed in the hard state of GX 339--4 by \citet{m08}. However, they were obtained using the non-relativistic Comptonization model of \citet{st80}, which is not valid for spectra extending to $\ga 50$ keV.}. In order to see if the 50--100 keV range can be extended to lower values, we have analyzed the \xte\/ data for a luminous hard state of the black-hole candidate GX 339--4 (from 2002 April 23, the observation ID 70109-01-04-00), with the bolometric accretion luminosity of $L_{\rm X}\simeq 2.4\times 10^{38}$ erg s$^{-1}$ at the assumed distance of 8 kpc \citep{z04}, and we have found using {\tt compps} $kT_{\rm h}\simeq 33_{-2}^{+3}$ keV (with $\chi^2_\nu=82/128$ including a systematic error of the data of 2 per cent). On the other hand, the values of $L_{\rm X}$ implied by our model (at $d=9$ kpc and assuming isotropic emission) in both of the analyzed hard states of Cyg X-3 are $\simeq 1.2\times 10^{38}$ erg s$^{-1}$, relatively similar to that in GX 339--4. Thus, $kT_{\rm h}\simeq 30$ keV in Cyg X-3 appears to be an entirely plausible value.

The temporal Green's function for $\tau_0=7.5$ (which is almost the same at $kT_0\ll 1$ keV as at $kT_0\simeq 3$ keV) is approximated by equation (\ref{fit}) and it is shown in Fig.\ \ref{tau5_7.5}(b). Its squared transform, $\left|{\cal G}(\phi)\right|^2$, is shown in Fig.\ \ref{attenuation}. We have fitted it to the observed power spectrum at high frequencies (assuming the intrinsic power $\propto f^{-1}$), as shown in Fig.\ \ref{power}. The fit implies $R\simeq 1.8\times 10^9$ cm, almost the same as given by equation (\ref{size}). It is somewhat smaller than that found in Section \ref{s:timing} due to the lower value of $\tau_0\simeq 5$ found in the models neglecting cloud heating.

\section{Discussion}
\label{s:discussion}

Given the obtained small size of the scattering cloud of $\sim\!2\times 10^9$ cm, it cannot be the stellar wind itself. A likely origin of the cloud appears to be collision of the gravitationally focused stellar wind with the outer edge of an accretion disc. The resulting cloud or bulge will be axially asymmetric with respect to the compact object (as required by the observed orbital modulation of X-rays). The presence of such a bulge was recently found in the high-mass black-hole binary Cyg X-1 \citep*{pzi08}. In Cyg X-1, the bulge is Thomson thin. However, the orbital separation, $a$, in Cyg X-3 is about 10 times lower than that in Cyg X-1, as well as the mass loss rate is at least 10 times higher (e.g., SZ08). Thus, the bulge in Cyg X-3 can easily have $\tau_0\sim 7$ and engulf the X-ray source. 

The size of the bulge is then similar to that of the disc. Since accretion in Cyg X-3 proceeds via stellar wind, its outer radius, $R_{\rm d}$, is much smaller than that of the Roche lobe \citep{sl76}. Using their formalism, $R_{\rm d}\simeq 4 a b^{-8} (1+M_2/M_{\rm X})^{-3}$, where $b=v_{\rm rel}/v_{\rm orb}$, $v_{\rm orb}$ is the relative orbital velocity, $v_{\rm rel}^2=v_{\rm orb}^2+v_{\rm wind}^2$, and $v_{\rm wind}$ is the wind velocity, which is $\simeq 10^8$ cm s$^{-1}$ (e.g., SZ08). Then, $R_{\rm d}\sim R$, e.g., $R_{\rm d}\simeq 2\times 10^9$ cm for plausible values of $M=30\msun$ and $M_2/M_{\rm X}=3$ \citep{v09}. This also yields \citep{sl76} the fraction of the mass loss rate captured by the compact object in a rough agreement with our value of $L_{\rm X}$. 

The characteristic ionization parameter of the cloud is $\xi\sim L_{\rm X}/(n_{\rm e} R^2)\sim 10^4$ erg cm s$^{-1}$, which implies that the cloud will be almost fully ionized. At the obtained rather high equilibrium $kT_0\simeq 3$ keV, the plasma will also be strongly collisionally ionized. The observed Fe K line and edge are thus formed mostly outside the cloud, in the stellar wind (e.g., SZ08). Still, there may be some contribution of the scattering cloud to absorption, which may be tested with high energy-resolution data, e.g., from {\it Chandra}.

For the sake of simplicity, we have assumed the scattering cloud to have a uniform density. In reality, it will have a density gradient and no sharp edge. An effect of that will be a less gradual cut-off of the attenuation function. However, this effect is most important for low values of $\tau_0$. At $\tau_0\simeq 7$, most of the scattering takes place relatively deep in the cloud and the surface effect is minor. 

An important finding of this study is the importance of thermal bremsstrahlung, which is otherwise found to be rather unimportant in black-hole binaries. In the present case, its importance is related to the presence of a dense and compact scattering cloud, apparently due to a high density of the stellar wind at a very high mass loss rate in this very compact binary. The luminosity of the bremsstrahlung turns out to be just at the right level to yield the energy balance of the cloud, which is highly remarkable. As we see from equation (\ref{brems}), $L_{\rm b}$ is determined by the cross section of the process, $\tau_0$, $kT_0$ and $R$, and it could be in principle at any level. Still, the values of $\tau_0$ and $kT_0$ resulting from the spectral fitting and the value of $R$ determined by the observed power spectrum result in the value of $L_{\rm b}$ just right to compensate for down-scattering losses of the primary emission due to thermal Comptonization in a hot plasma. We consider this coincidence a strong argument for the correctness of the presented model. We note that an inhomogeneity of the scattering cloud would further enhance the bremsstrahlung emission by a factor of $q$, see equation (\ref{f}).

It is also of interest to consider the dynamical state of scattering cloud. It cannot be free-falling as it would have given a too large accretion rate. It may be partly pressure supported. The dominant contribution to the pressure, $P$, is found to be radiation, $P\sim \tau_0^2 L/(4\upi R^2 c)$. Then, the hydrostatic equilibrium, $P/R\sim GM_{\rm X} m_{\rm p} n_{\rm e} \mu_{\rm e}/R^2$ yields $M_{\rm X}/\msun\sim \tau_0 L/L_{\rm E}^1$, where $L_{\rm E}^1$ is the Eddington luminosity for $1\msun$ (which is higher for He than that for H by the factor of $\mu_{\rm e}\simeq 2$). This yields $M_{\rm X}$ of several $\msun$. This is, however, only an order of magnitude determination given our idealized model of a uniform cloud. Furthermore, the cloud is likely to be in a state of sub-Keplerian rotation, and thus partly centrifugally supported, which would increase the required $M_{\rm X}$. Thus, our model is consistent with the presence of a black hole in Cyg X-3. 

On the other hand, we note that our determination of the size of the cloud rests on the cut-off in the power spectrum. If that cut-off has a different origin, the scattering cloud could have a much larger size, and thus be effectively cooled by the stellar photons. Then, the models presented in Section \ref{downscattering}, which are not consistent with the derived small size, could still be the correct ones. We note, however, that those models yield the temperatures of the intrinsic hot plasma substantially lower (and thus appearing less realistic) than both those of the models in Section \ref{s:equilibrium} and those in the hard state of black-hole binaries.

Interestingly, our calculations yield almost the same intrinsic luminosities, $L_{\rm X}$, for the two hard spectral states considered. In general, we expect softer states to be more luminous (e.g., \citealt*{dgk07}). However, hysteresis effects in black-hole binaries often lead to different states to occur at the same luminosity (e.g., \citealt{z04}), and there are indication that such an effect also occurs in Cyg X-3 \citep{sv96}. On the other hand, we have modelled only the hard part of the X-ray spectra. Given the very strong absorption in this system, there may be an additional soft component, stronger in the softer of the two considered states. An indication for that is the difference in the fitted values of $N_{\rm H}$, with the apparently lower value of it in the softer state being possibly due to the presence of an additional soft emission.

Our obtained value of $L_{\rm X}\simeq 1.2\times 10^{38}$ erg s$^{-1}$ is $\sim\! 50$--70 per cent of the values obtained for the corresponding spectral states by Hj09. Based on a comparison of the colour-luminosity diagram for Cyg X-3 with that for black-hole binaries, Hj09 favoured $M_{\rm X}\simeq\! 30\msun$. Our lower value of $L_{\rm X}$ would imply a mass of $\sim\! 2/3$ of that, provided this argument holds. Also, the parameters of the intrinsic X-ray emitting plasma derived by Hj09 would need to be significantly revised due to the existence of a scattering cloud postulated in our model.

\begin{figure}
\centerline{\includegraphics[width=7.cm]{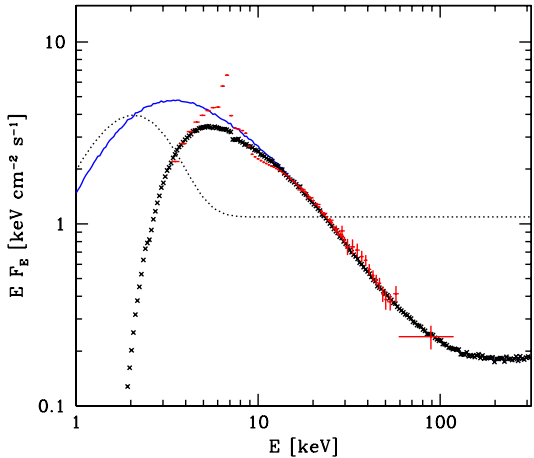}} 
\caption{A preliminary comparison of the spectrum of the average state 3 (soft) of Cyg X-3, shown by the error bars, with a model consisting of an intrinsic emission of a Wien spectrum and a power law (dots) passing through a scattering cloud with the same parameters as those found for the state 2, $\tau_0=7.0$, $kT_0=2.5$ keV. In the intrinsic spectrum, a power law with the photon index of 2 is convolved with a Wien spectrum with the temperature of 0.5 keV. The solid curve shows the intrinsic spectrum after transmission through the cloud, and the crosses show the transmitted spectrum after absorption by external matter with $N_{\rm H}=7\times 10^{22}$ cm$^{-2}$. For simplicity, neither bremsstrahlung nor Compton reflection/Fe K emission are taken into account.
}
\label{state3}
\end{figure}

An important issue is whether our model is applicable also to soft states of Cyg X-3 (3, 4, 5 in the classification of SZM08). Since Compton scattering of photons at a given energy moves them to a wide range of energies, there is no unique intrinsic spectrum corresponding to a given observed spectrum for given parameters of the scattering cloud, $kT_0$ and $\tau_0$. Furthermore, the parameters of the cloud may change between states, in particular $kT_0$ follows from energy balance and thus depends on the incident spectrum. Then, a possible approach is to take observed soft state spectra of black-hole binaries and see how Compton scattering modifies them. However, this is a highly complex problem, given the wide range of observed soft state spectra, which requires a relatively large number of free parameters to model (e.g., \citealt{g99,zg04,dgk07}). Thus, we defer its systematic study to future work. Here, we just consider one spectrum, which we model in a very simple way. We take the average spectrum 3 and the cloud parameters fitted to the spectrum 2. As a simple intrinsic spectrum, we use the {\tt bmc} model of {\sc xspec}, which is a convolution of a power law with a Wien spectrum\footnote{We note that although {\tt bmc} is intended to model bulk-motion Comptonization in a converging flow around a black hole, it does not due to not including a high energy cut-off, which is always at $\la\! 100$ keV for that process \citep{nz06}. Still, {\tt bmc} is a useful generic approximation to Comptonization by non-thermal electrons.}. We find that this model gives a relatively good description of the average spectrum 3, see Fig.\ \ref{state3}. Since we use a phenomenological, i.e., non-physical description of the input spectrum, we do not consider here the issues of energy balance or bremsstrahlung emission (which we defer to a future study with a physical model, e.g., that of hybrid Comptonization, see, e.g., \citealt{g99}). Still, our results show the potential possibility of explaining also the soft spectra of Cyg X-3 by a model with transmission through a scattering cloud. 

A test of our model can be performed by studying \xte\/ data for the hard states binned with respect to the orbital phase. From the expected anisotropy of the scattering cloud, the optical depth along the line of sight should be lower around the superior conjunction (the compact object in front of the donor) than around the inferior one. This would result in the cut-off of both the energy and power spectra being higher for the former case than for the latter.

\section{Conclusions}
\label{conclusions}

We have considered the X-ray spectra of Cyg X-3 in its hard spectral state. The observed spectra show a cut-off/break at an unusually low energy. We find this can be explained if a standard X-ray source is surrounded by a Thomson-thick cold plasma, which down-scatters the intrinsic X-ray spectrum. The presence of such a plasma was earlier postulated to explain the energy-independent X-ray orbital modulation and the lack of high frequencies in the power spectra. 

We use a Monte Carlo code to model the effect of Compton scattering in the cold plasma on both spectra and variability. For the latter, we obtain Green's functions and their Fourier transforms. Our results show that Compton scattering affects the power spectra at a frequency $\sim\! 4$ times lower than that obtained from considering the average number of scattering in the cloud. 

Finally, we consider thermal equilibrium of the cloud in the field of the incident photons. The obtained model has the cloud temperature of $kT_0\simeq 3$ keV, the optical depth of $\tau_0\simeq 7$, and the cloud size of $R\simeq 2\times 10^9$ cm. We find that for those parameters, thermal bremsstrahlung emission of the cloud is important. Then, the cooling by this mechanisms facilitates achieving the exact energy equilibrium of the cloud. The model reproduces both the X-ray spectrum and the power spectrum in the hard state. The cloud is partly supported by the radiation pressure.

\section*{Acknowledgments}
We than C. Done for valuable discussions, K. Leszczy{\'n}ski for help with the Monte Carlo code, and the referee for inspiring suggestions. This research has been supported in part by the Polish MNiSW grants NN203065933 and 362/1/N-INTEGRAL/2008/09/0, and the Polish Astroparticle Network 621/E-78/BWSN-0068/2008. We acknowledge the use of data obtained through the HEASARC online service provided by NASA/GSFC.

\label{lastpage}

\end{document}